\documentclass{elsart}

\usepackage{graphicx}

\usepackage{amssymb}

\begin{document}
\begin{frontmatter}



\title{Perturbed phase-space dynamics of hard-disk fluids}


\author[a]{Christina Forster},  
\ead{tina@ap.univie.ac.at}
\author[a,b]{Robin Hirschl},  
\ead{Robin.Hirschl@univie.ac.at}
\author[a]{Harald A. Posch},
\ead{posch@ls.exp.univie.ac.at}
\author[c]{William G. Hoover}
\ead{hoover3@llnl.gov}
\address[a]{Institut f\"ur Experimentalphysik, Universit\"at Wien,
Boltzmanngasse 5, A-1090 Vienna, Austria}
\address[b]{Institut f\"ur Materialphysik, Universit\"at Wien,
Boltzmanngasse 5, A-1090 Vienna, Austria}
\address[c]{Department of Applied Science, University of California at
        Davis/Livermore and Methods Development Group, Lawrence Livermore
        National Laboratory, Livermore, California 94551-7808, USA}

\begin{abstract}
    The Lyapunov spectrum describes the exponential growth, or decay, of
infinitesimal phase-space perturbations.  The perturbation associated with 
the maximum Lyapunov exponent is strongly localized in space, and only a small 
fraction of all particles contributes to the perturbation growth at any 
instant of time.  This fraction converges to zero in the 
thermodynamic large-particle-number limit.  For hard-disk and hard-sphere
systems the perturbations belonging to the small positive and large negative 
exponents are coherently spread out and form orthogonal periodic structures 
in space, the ``Lyapunov modes''. There are two types of mode polarizations,
transverse and longitudinal. The transverse modes do not propagate, but the 
longitudinal modes do with a speed about one third of the sound speed. We 
characterize the symmetry and the degeneracy of the modes. In the
thermodynamic limit the Lyapunov spectrum has a diverging slope
near the intersection with the abscissa. No positive lower bound
exists for the positive exponents. The mode amplitude scales with
the inverse square root of the particle number as expected from the
normalization of the perturbation vectors.
\end{abstract}

\end{frontmatter}

\section{Introduction}
\label{Sec1}

Recently, the concepts of dynamical systems theory and
molecular dynamics simulations have been applied to the
phase-space dynamics of models representing simple fluids and
solids \cite{ph88,PH89}.  Due to the convex and dispersive nature of the 
atomic surfaces, the phase trajectory of such a system is highly Lyapunov 
unstable resulting in exponential growth, or decay, of small (infinitesimal) 
initial perturbations along specified directions in phase space.
The respective rate constants, the 
Lyapunov exponents, are taken to be ordered by size with $\lambda_1>0$ 
representing the maximum exponent. The whole set, 
$\{\lambda_l,  l=1,\ldots,2dN\} $, is known as the Lyapunov spectrum, 
where $d$ denotes the dimension of space, and $N$ is the number of particles.
For fluids in nonequilibrium steady states close links exist between the
Lyapunov spectrum and various macroscopic dynamical properties -- transport 
coefficients, irreversible entropy production, and the Second Law of 
thermodynamics \cite{ph88,PH89,HOO,em90,GASP,DO99}. This important result
provides the motivation for us to examine also the spatial structure of the 
perturbations associated with the various exponents. Here we restrict
ourselves to systems in thermodynamic equilibrium.  Earlier accounts of
our work have been published  in Refs.
\cite{ph00,h99,pf02,hpfdz,mph98a,mph98b,mp02,m01}.
 
    Since the pioneering work of Bernal \cite{BERNAL} with steel balls,
and the seminal computer simulations of Alder and coworkers \cite{b3,EIN},
hard ball systems are considered good models for the structure of
``real'' dense fluids.  They serve as reference systems for highly-successful 
perturbation theories of liquids \cite{Hansen,Reed}. The dynamics, however,
is rather different, and it is rewarding to study this difference, not
only with the established machinery of correlation functions, but
also from the viewpoint of dynamical systems theory. 
In this paper we discuss our results on the Lyapunov spectra and the
space dependence of the associated perturbation vectors for two-dimensional
hard disk systems in equilibrium. For comparison we also present
related results for two-dimensional soft-disk fluids.

\begin{figure}
\centering{\includegraphics[width=10cm,clip=, angle=-90]{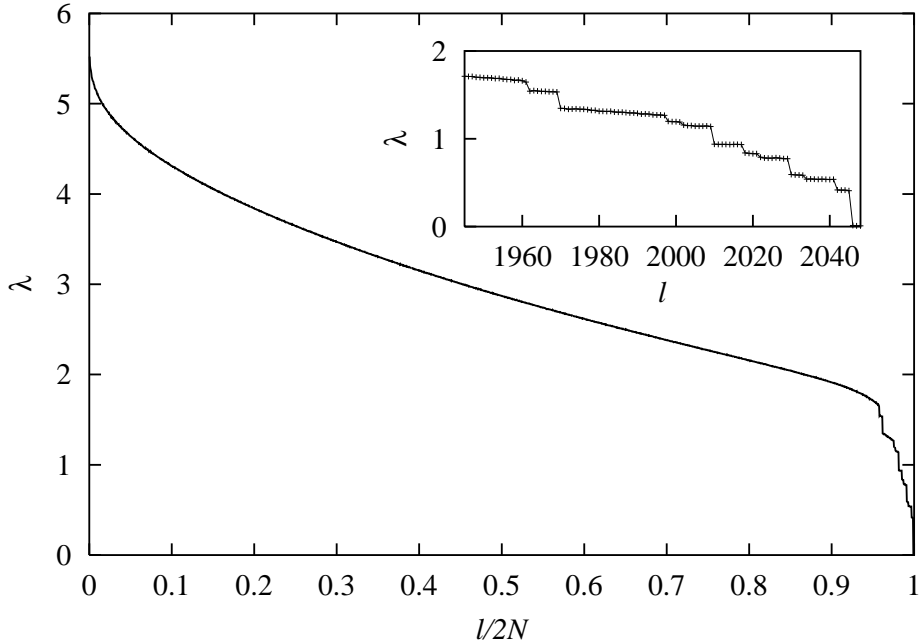}}
\caption{Lyapunov spectrum of a 1024-disk system in a square simulation box
with periodic boundaries. The density $\rho = 0.7$. A reduced index
$l/2N$ is used on the abscissa. The steps are magnified in the inset. 
The non-normalized index, $l$, is used there. The spectrum is defined only
for integer values of $l$. }
\label{Fig1}
\end{figure}

    This work has three main objectives: \\ \noindent
1) In Section \ref{Sec3} we demonstrate that the perturbation associated
with the maximum Lyapunov exponent, $\lambda_1$, is strongly localized in 
space:  only a small fraction of all particles contributes actively 
to the growth of the perturbation norm at any instant of time. 
This localization persists in the thermodynamic limit, and, even more 
surprisingly, the fraction of simultaneously-contributing degrees of freedom
converges to zero. 
\\ \noindent
2) For small exponents close to the intersection with the abscissa the
 Lyapunov spectrum has a step-like appearance as is depicted in Fig.
\ref{Fig1} for a two-dimensional 1024-disk fluid with particle density 
$\rho = 0.7$. 
The steps are caused by degenerate exponents, and the associated perturbations 
form coherent wave-like patterns in space with well-defined wave vectors and 
polarization.  We  refer to these patterns as ``Lyapunov modes''.
We have found them in hard-disk and hard-sphere systems in two and 
three dimensions \cite{ph00,h99,pf02,hpfdz}, respectively,
and in hard-dumbbell systems modelling fluids consisting of linear
molecules \cite{mph98a,mph98b,mp02,m01}.  In Section \ref{Sec4} we 
characterize the symmetry and polarization of these modes, where the emphasis 
is again on the thermodynamic limit for bulk systems: $N$ and $V \to \infty$ 
such that $\rho =$const.  $V$ is the area of the box. We show that in 
this limit the Lyapunov spectrum has a vertical slope
at the intersection point with the abscissa.  No gap, or lower positive 
bound, exists for the positive exponents, and the smallest positive exponent 
converges linearly to zero with inverse system size. 
\\ \noindent
3) How do the modes depend on the intermolecular potential?  In Section 
\ref{Sec6}
we discuss simulations of two-dimensional particle systems 
interacting with a continuous repulsive Weeks-Chandler-Anderson (WCA) 
potential. They reveal that the modes are extremely unstable and not easily 
observed. We relate this phenomenon to the strong fluctuations of the
time-dependent local Lyapunov exponents.

Since our discovery of the Lyapunov modes the theoretical foundation of the 
Lyapunov modes has been examined in some detail
\cite{Eckmann,Zabey,McNM01,MMcN02,deWvB,TDM02,TM02}.
This work will be summarized in the concluding Section \ref{Sec7}.

To facilitate the interpretation of our numerical results we outline our 
numerical methods for the computation of Lyapunov spectra for hard-disk 
systems \cite{DPH96,ph00} in the following section.

\section{Lyapunov spectra and numerical procedure}
\label{Sec2}

The instantaneous  state of a planar particle system is given by the
$4N$-dimensional phase-space vector  $ {\bf \Gamma} = \{{\bf q}_i, {\bf p}_i,
; i = 1,\ldots,N \}$, where ${\bf q}_i= (x_i, y_i)$ and 
${\bf p}_i = (p_{x,i}, p_{y,i})$ denote the
position and linear momentum of particle $i$, respectively. An infinitesimal
perturbation or offset vector 
$ \delta {\bf \Gamma} = \{\delta{\bf q}_i, \delta {\bf p}_i;
i = 1,\dots,N \}$ evolves according to motion equations obtained by
linearizing the dynamical equations for ${\bf \Gamma}(t)$.  For hard disks
this amounts to a linearization of the collision map, which maps pre-collision 
into post-collision states in phase space \cite{DPH96}. 
There exist $4N$ orthonormal 
initial vectors  $ \{ \delta {\bf \Gamma}_l(0); l = 1,\ldots ,4N\}$
in tangent space, such that the Lyapunov exponents 
\begin{equation}
  \lambda_l = \lim_{t\to\infty}\frac{1}{t} \ln
              \frac{|\delta {\bf\Gamma}_l(t)|}{|\delta {\bf\Gamma}_l(0)|}
              \;\; , \;\;l = 1,\ldots ,4N \;  ,
\label{lyap}
\end{equation}
exist and are independent of the initial state \cite{oseledec,eckmann}. 
Geometrically, the Lyapunov spectrum describes the stretching and contraction 
along linearly-in\-de\-pen\-dent phase space directions of an infinitesimal 
hypersphere co-moving with the flow. According to the conjugate pairing rule 
\cite{DM96,Ruelle,ECM90} for symplectic systems, the exponents 
appear in pairs, and the 
pair sum vanishes, $\lambda_{l} + \lambda_{4N+1 - l} = 0$.
Thus, only the positive half of the spectrum $\{\lambda_{1 \le l \le 2N}\}$
needs to be calculated.  Furthermore, six of the exponents,
$\{\lambda_{2N-2 \le l \le 2N+3}\}$,  vanish as a consequence
of the conservation of energy, momentum, and center of mass, and of
the non-exponential time evolution of a perturbation vector parallel to the
phase flow.
   
For our numerical work we use a variant \cite{DPH96,ph00} of the 
classical algorithm by Benettin 
{\em et al.}  and Shimada and Nagashima \cite{Benettin,Shimada}, where the
reference trajectory and the orthonormal set of perturbation vectors
are simultaneously evolved in time. The latter are
periodically re-orthonormalized with a Gram-Schmidt (GS) procedure after
consecutive  time intervals $\Delta t_{GS}$, which were chosen according to
$\Delta t_{GS} \simeq \tau_2 /5 $. Here,
$\tau_2 = (\beta m)^{1/2} / (2 \pi^{1/2} \rho \sigma) $ is
the collision time per particle, $\beta = (k_B T)^{-1}$, 
$T$ is the temperature, $k_B$ is Boltzmann's constant, and $\sigma$ is the
disk diameter.  The Lyapunov exponents are determined from the time-averaged
renormalization factors.  

Fig. \ref{Fig1} shows the positive branch of the Lyapunov spectrum for a 
1024-disk system in a square simulation box with periodic boundaries.
If not stated otherwise, reduced units are used for which the particle 
diameter $\sigma$, the particle mass $m$, and the kinetic energy per particle, 
$K/N\equiv k_B T$, are unity.
Thus, the unit of time is $t^* = (m \sigma^2N / K)^{1/2}$, 
and the Lyapunov exponents are given in units of $1/t^*$. 
The simulation box has lengths $L_x$ and $L_y$. In all our examples
the density $\rho \equiv N/(L_x L_y)$ is 0.7, which is slightly smaller than the
fluid-to-solid phase transition density.

\section{Localized mode for the maximum exponent}
\label{Sec3}

The maximum exponent is dominated by the fastest dynamical events.  There is 
strong numerical evidence that the thermodynamic limit $\{ N,V \to \infty, 
N/V$con\-stant$\}$ for $\lambda_1$ and, hence, for the whole spectrum exists 
\cite{DPH96,mp02}.  The associated perturbation is strongly localized
in space \cite{HBP98,mph98b,mp02}. This is demonstrated by
projecting the perturbation vector $\delta {\bf \Gamma}_1$
onto the four-dimensional subspaces spanned by the components
belonging to the individual particles.  The squared norm of this projection,
$|\delta {\bf q}_i|^2 + |\delta {\bf p}_i|^2$,
indicates how active a particle $i$ is engaged in the growth process
characterized by $\lambda_1$ \cite{mp02,mph98b}.
\begin{figure}
\centering{\includegraphics[width=8cm,clip=,angle=-90]{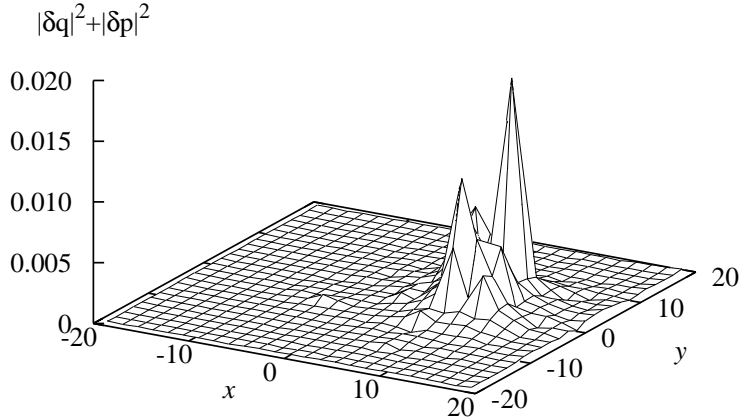}}
\vspace{-5mm}
\caption{Squared norm of the individual particle perturbations, 
$|\delta {\bf q}_i|^2 + |\delta {\bf p}_i|^2$, for the offset vector
associated with $\lambda_1$, plotted at the positions ${\bf q}_i = (x_i,y_i)$
of the particles.
The ensuing surface is linearly interpolated over a periodic grid covering the
simulation box.  The system consists of 1024 hard disks in a square box with 
periodic boundaries, at a density $\rho = 0.7$, and a temperature $T = K/N = 1$.} 
\label{Fig2}
\end{figure}
In Fig. \ref{Fig2} it is plotted for a 1024-disk system at the 
particle positions $(x_i,y_i)$, where the ensuing surface is interpolated over 
a periodic grid covering the simulation box. The figure refers to an 
instantaneous configuration of a well-relaxed system.  The fastest 
expansion activity is clearly confined to a few very active zones.

   To understand this localization we note that the offset-vector dynamics is 
governed by {\em linear} equations and that a perturbation component has 
only a fair chance of further growth if its value was already above average 
before a collision. Each global renormalization step tends to reduce 
the (already small) components of the other non-colliding particles even 
further. There is a competition for growth which favors the particles with the
largest perturbation components. The few active zones move diffusively
in space and perform occasional jumps.

   The localization persists in the thermodynamic limit. To
show this \cite{mp02} we order
all squared components $[\delta{\bf\Gamma}_l]^2_j; j = 1,\dots,4N$
of a perturbation vector $\delta {\bf \Gamma}_l$ according to size.
By adding them up, starting with the largest, we determine
the smallest number of terms, $n$, which are required for the sum
to exceed a threshold $\Theta$.   $ C_{l,\Theta} \equiv n/4N$
is a {\em relative} measure
for the number of components contributing to $\lambda_l$:
\begin{equation}
        \Theta \le\left\langle \sum_{s=1}^{4N C_{l,\Theta}}
        \left[\delta{\bf\Gamma}_l\right]^2_s\right\rangle,
        \;\;\;\;\;\;\;\;
        [\delta{\bf\Gamma}_l]^2_i \ge [\delta{\bf\Gamma}_l]^2_j \;{\rm
        for}\; i<j.
\end{equation}
\begin{figure}
\centering{\includegraphics[width=8cm,clip=, angle=-90]{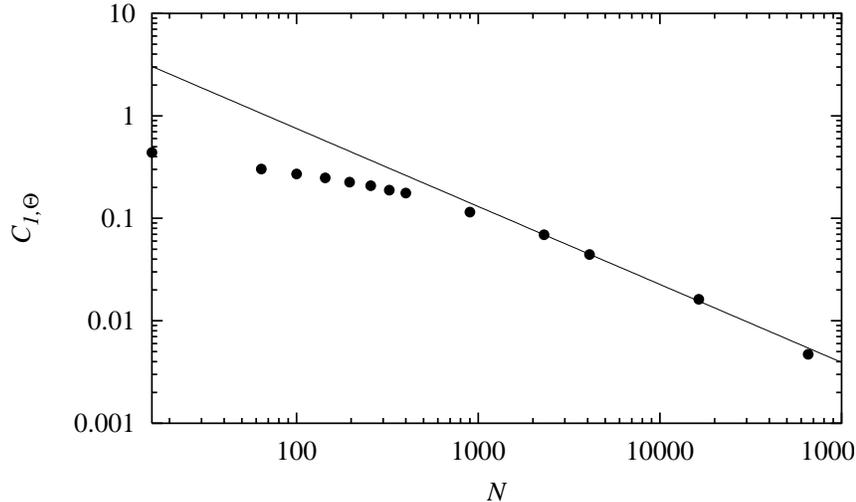}}
\caption{Particle-number dependence of the localization measure 
for the perturbation belonging to the maximum exponent, $C_{1,\Theta}$. 
The threshold $\Theta$ is 0.98. The line represents a power law,
$25 N^{-0.76}$.}
\label{Fig3}
\end{figure}
Obviously, $C_{l,1}=1$.  In Fig. \ref{Fig3} the localization measure, for 
$l=1$ and for a threshold $\Theta = 0.98$, is shown as a function of the 
particle number $N$. It obeys a power law with a negative power, and 
converges to zero for $N\to\infty$.  The ratio of tangent-vector components 
(and, hence, of particles) contributing significantly to the
maximum Lyapunov exponent vanishes in that limit. 
A power-law dependence for the localization measure is a very general result, 
although the parameters are sensitive to the interaction potential 
\cite{pf02,fp03}.  

\begin{figure}
\centering{\includegraphics[width=10cm,clip=,angle=-90]{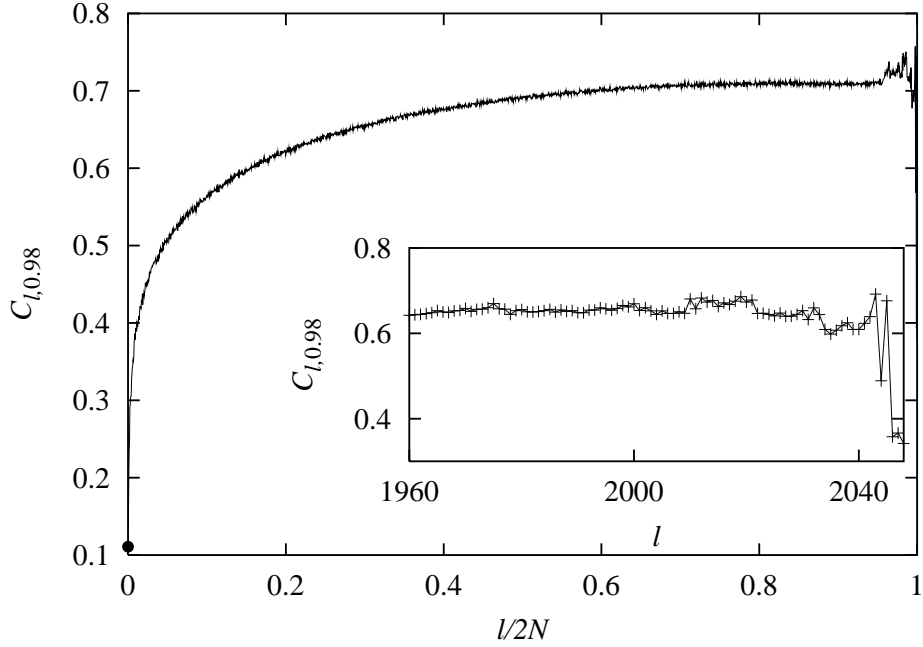}}
\caption{Dependence of the localization measure $C_{l,\Theta = 0.98}$ 
on the reduced Lyapunov index $l/2N$. The system consists of
$N=1024$ particles, and the density $\rho$ is 0.7.  The inset
provides an enlargement of the mode regime. Here, the Lyapunov index
$l$ is not normalized.}
\label{Fig4}
\end{figure}

If the localization measure is plotted for all positive exponents
with index $1 \le l \le 2N$ as depicted in Fig. \ref{Fig4}, 
one observes a fast increase for small $l$, where the full dot indicates the
result for $l=1$. This is an indication that
the localization does not persist for large $l$ and that the perturbation 
growth starts to involve more and more particle components.
The inset shows the interesting regime for which modes have been observed.

\section{Perturbation vectors for the six vanishing exponents}
\label{Sec4}
  
    To be explicit, we arrange the components of the $4N$-dimensional
state vector according to
\begin{eqnarray}
{\bf \Gamma} = (x_1,y_1,\dots,x_N,y_N;p_{x,1},p_{y,1},\dots,p_{x,N},p_{y,N}) \;.
\end{eqnarray}
During the streaming motion in between instantaneous elastic collisions
no forces act between the particles, and a normalized vector parallel to the 
phase flow $\dot{\bf \Gamma}$ becomes 
\begin{eqnarray}
{\bf e}_1 = (1/\sqrt{2K}) (p_{x,1},p_{y,1}\dots,p_{x,N},p_{y,N};
                    0,0,\dots,0,0)\; .
\label{e1}
\end{eqnarray}
A perturbation vector parallel to ${\bf e}_1$ does not
grow exponentially with time and contributes one vanishing exponent to the
full spectrum \cite{GASP}.

  The motion in the extended $4N$-dimensional phase space is constrained
to hypersurfaces generated by center-of-mass conservation,
($\sum x_{j=1}^N =$ const, $\sum y_{j=1}^N $ = const), by momentum conservation,
($\sum_{j=1}^N p_{x,j} = 0$, $\sum_{j=1}^N p_{y,j} = 0$), and by the 
conservation of
energy ($\sum_{j=1}^N (p_{x,j}^2 + p_{y,j}^2 )/2 = K$). All respective 
vectors orthogonal to these surfaces also denote directions of 
non-exponential phase-space expansion or compression:
\begin{eqnarray}
{\bf e}_2 &=& (1/\sqrt{N}) (1,0,\dots,1,0;0,0,\dots,0,0) \label{e2}\\
{\bf e}_3 &=& (1/\sqrt{N}) (0,1,\dots,0,1;0,0,\dots,0,0) \label{e3}\\
{\bf e}_4 &=& (1/\sqrt{N}) (0,0,\dots,0,0;1,0,\dots,1,0) \label{e4}\\
{\bf e}_5 &=& (1/\sqrt{N}) (0,0,\dots,0,0;0,1,\dots,0,1) \label{e5}\\
{\bf e}_6 &=& (1/\sqrt{2K})(0,0,\dots,0,0;p_{x,1},p_{y,1},
                                \dots,p_{x,N},p_{y,N}) \;.\label{e6}
\end{eqnarray}
The  vectors ${\bf e}_1$  to ${\bf e}_6$ are orthonormal and form the
basis of an invariant subspace, the central manifold, which contains also 
the six perturbation vectors $\{\delta{\bf \Gamma}_l$,  
$ 2N-2\le l \le 2n+3\}$ responsible for the vanishing
exponents. An expansion in the natural basis gives
\begin{equation}
\delta{\bf \Gamma}_l  = \sum_{i=1}^6 \alpha_{l,i} \; {\bf e}_i \;, \; l =
      2N-2, \dots, 2N+3 \; .
\end{equation} 
The matrix of the projection cosines, 
$\alpha_{l,i} = (\delta{\bf \Gamma}_l \cdot {\bf e}_i)  $,
obeys the normalization conditions
$$\sum_{i=1}^6 (\alpha_{l,i})^2 = 1\;\;\;, \;\;\;  
  \sum_{l=2N-2}^{2N+3}  (\alpha_{l,i})^2 = 1 \; ,$$  
if the total momentum vanishes and if 
$K = N$ as is always the case in our simulations.  As an example we list such a 
matrix in Table \ref{table1} for a 400-disk fluid. We note that these numbers
\begin{table}
\caption{Projection cosines  of the perturbation vectors in the 
central manifold onto the natural basis vectors for a well-relaxed 
configuration of a 400-disk fluid with density $\rho =0.7$ and a total 
kinetic energy $K = N$.  }
\label{table1}
\begin{center}
\begin{tabular}{c|rrrrrr} \hline \hline
 $l$ & $ \alpha_{l,1} $ & $\alpha_{l,2} $ & $\alpha_{l,3} $ & $\alpha_{l,4} 
$ & $\alpha_{l,5} $ & $ \alpha_{l,6}   $  \\ \hline

 398  & 0.431 &  0.297  & -0.852  &  0 &  0 & 0  \\[-2mm]
 399  & 0.549 & -0.836  & -0.013  &  0 &  0 & 0  \\[-2mm]
 400  &-0.716 & -0.462  & -0.523  &  0 &  0 & 0  \\[-2mm]
 401  & 0 & 0  & 0  & -0.138 & -0.607 &-0.783    \\[-2mm]
 402  & 0 & 0  & 0  & -0.620 &  0.669 &-0.409    \\[-2mm]
 403  & 0 & 0  & 0  & -0.772 & -0.429 & 0.469 \\\hline    
\end{tabular} 
\end{center}
\end{table}
are the asymptotic values for a well-relaxed system and do not
change with time. They always appear in block-diagonal form, but 
other than that they depend on the 
initial conditions for the simulation. The block-diagonal structure
indicates that the subspaces 
$\mathcal{C}^+ = $span$\{{\bf e}_1, {\bf e}_2, {\bf e}_3\}$ and 
$\mathcal{C}^- = $span$\{{\bf e}_4, {\bf e}_5, {\bf e}_6\}$
are asymptotically ordered by the Gram-Schmidt procedure due to 
{\em linear} perturbation growth/decay during the
streaming motion between successive collisions. This may be seen
explicitly from the linear motion equations for intercollisional streaming
and the linearized collision map. For example, perturbations 
$\delta {\bf \Gamma}$, parallel to
${\bf e}_1$, ${\bf e}_2$, or  ${\bf e}_3$ at $t=0$ immediately
after a collision, will continue to be so at a later time $t>0$.
However, perturbations  $\delta {\bf \Gamma}$ parallel to 
${\bf e}_4$, ${\bf e}_5$, and  ${\bf e}_6$ at time zero, evolve
according to ${\bf e}_2 t + {\bf e}_4$,  ${\bf e}_3 t + {\bf e}_5$, 
and  ${\bf e}_1 t + {\bf e}_6$, respectively, for $t>0$ previous to the
next collision. 

\section{Lyapunov modes}
\label{Sec5}

\subsection{Classification of the modes}
    Next, we study the non-localized perturbations associated with 
the small positive exponents. The simulation box is a square with
$L_x = L_y = L$.  The inset in Fig. \ref{Fig1} gives an enlargement
of the step-like structure for a 1024-disk spectrum.

\begin{figure}
\centering{\includegraphics[width=5cm,clip=,angle=-90]{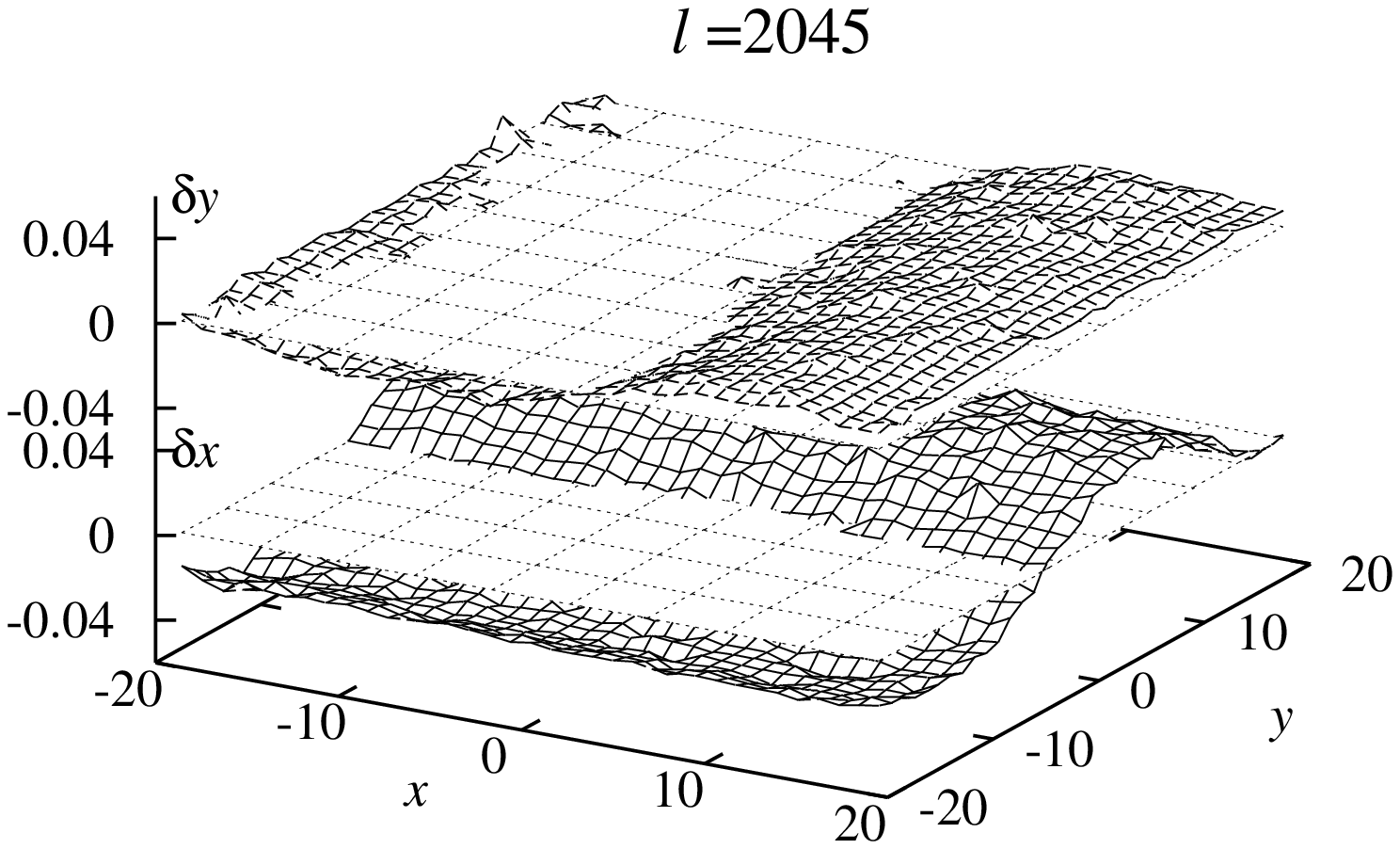}}
\hspace{1cm}
\centering{\includegraphics[width=5cm,clip=,angle=-90]{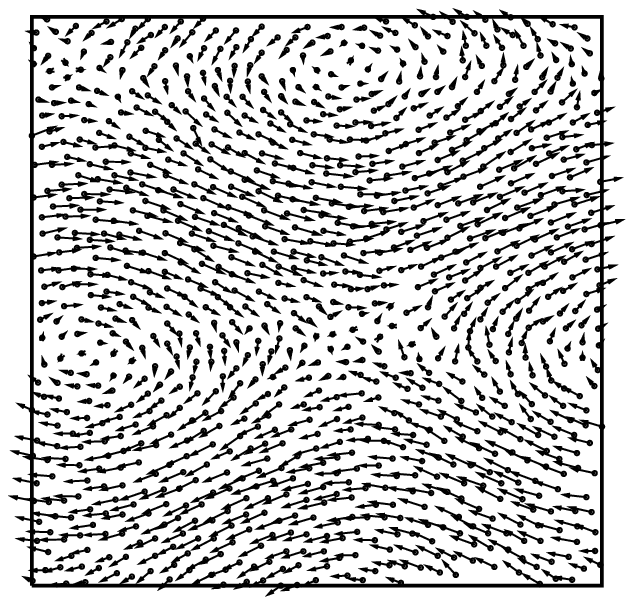}}
\centering{\includegraphics[width=5cm,clip=,angle=-90]{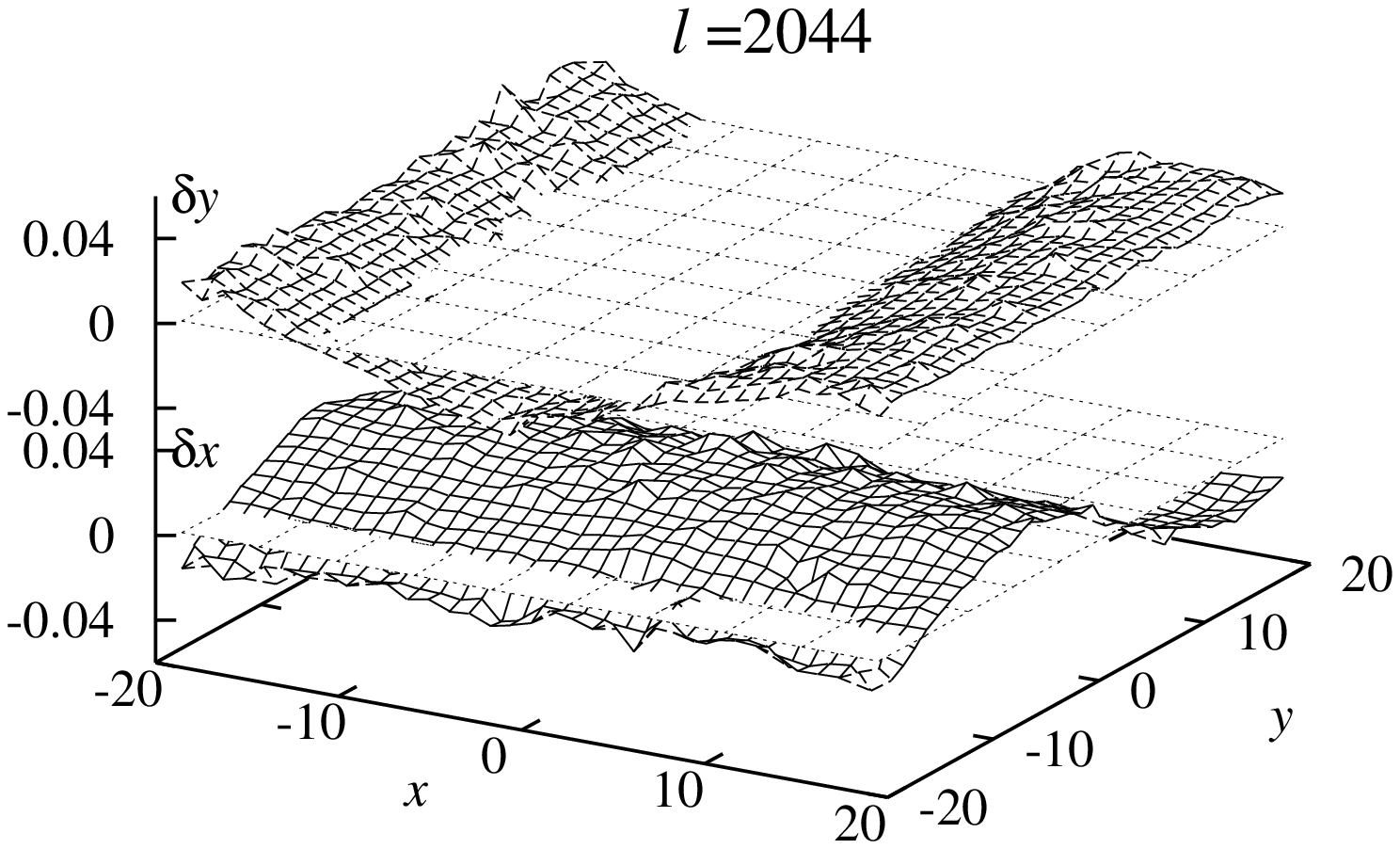}}
\hspace{1cm}
\centering{\includegraphics[width=5cm,clip=,angle=-90]{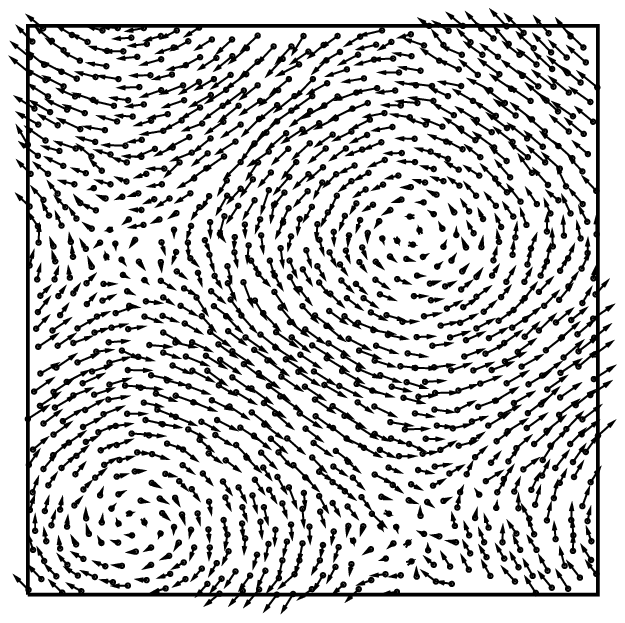}}
\centering{\includegraphics[width=5cm,clip=,angle=-90]{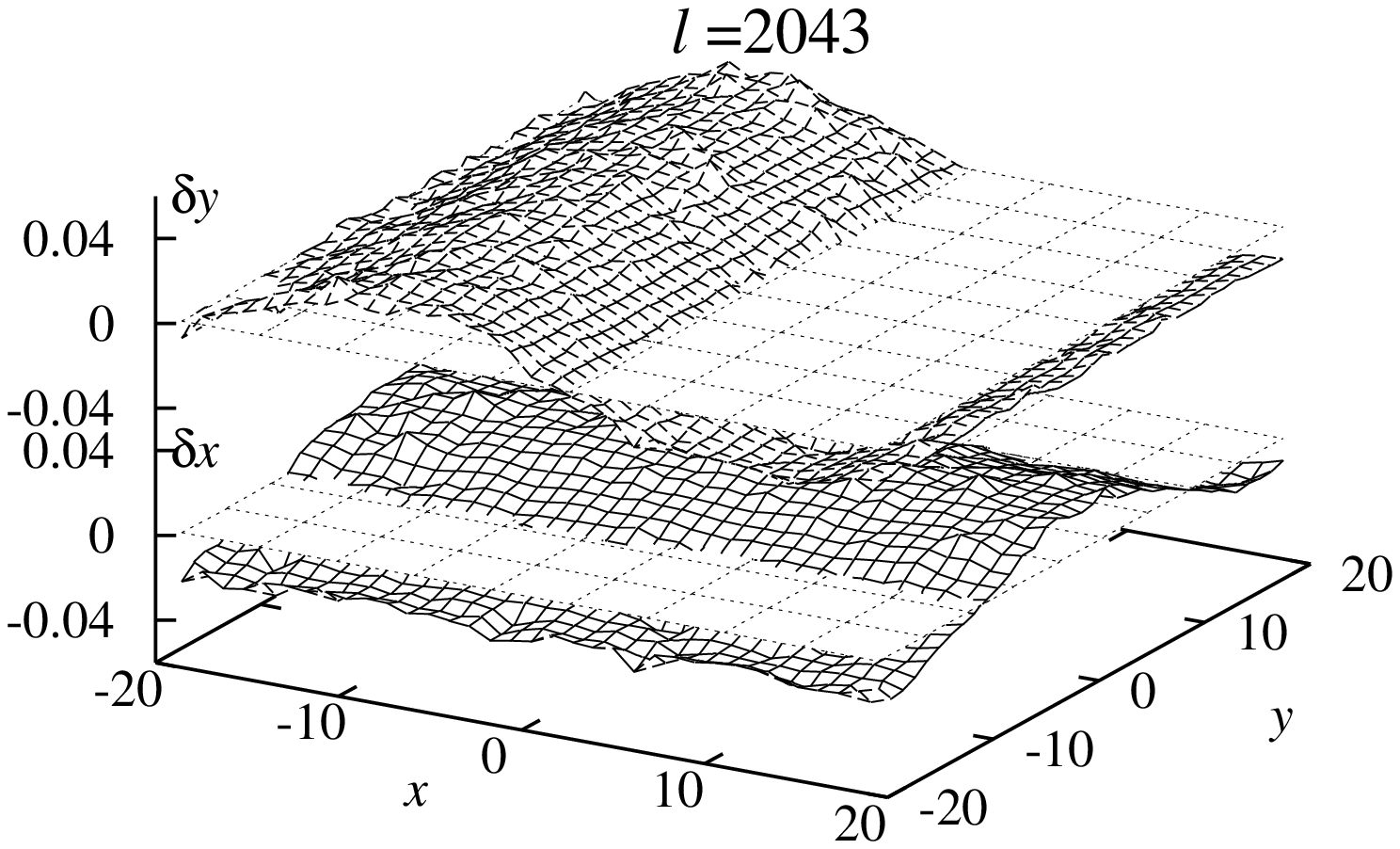}}
\hspace{1cm}
\centering{\includegraphics[width=5cm,clip=,angle=-90]{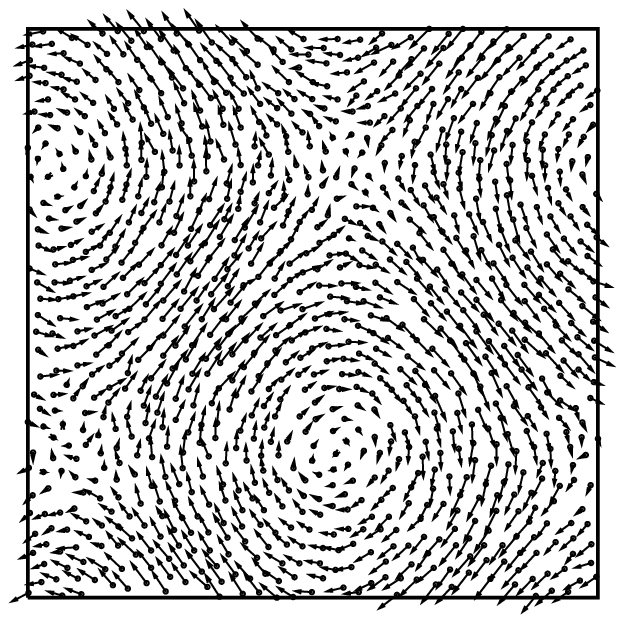}}
\centering{\includegraphics[width=5cm,clip=,angle=-90]{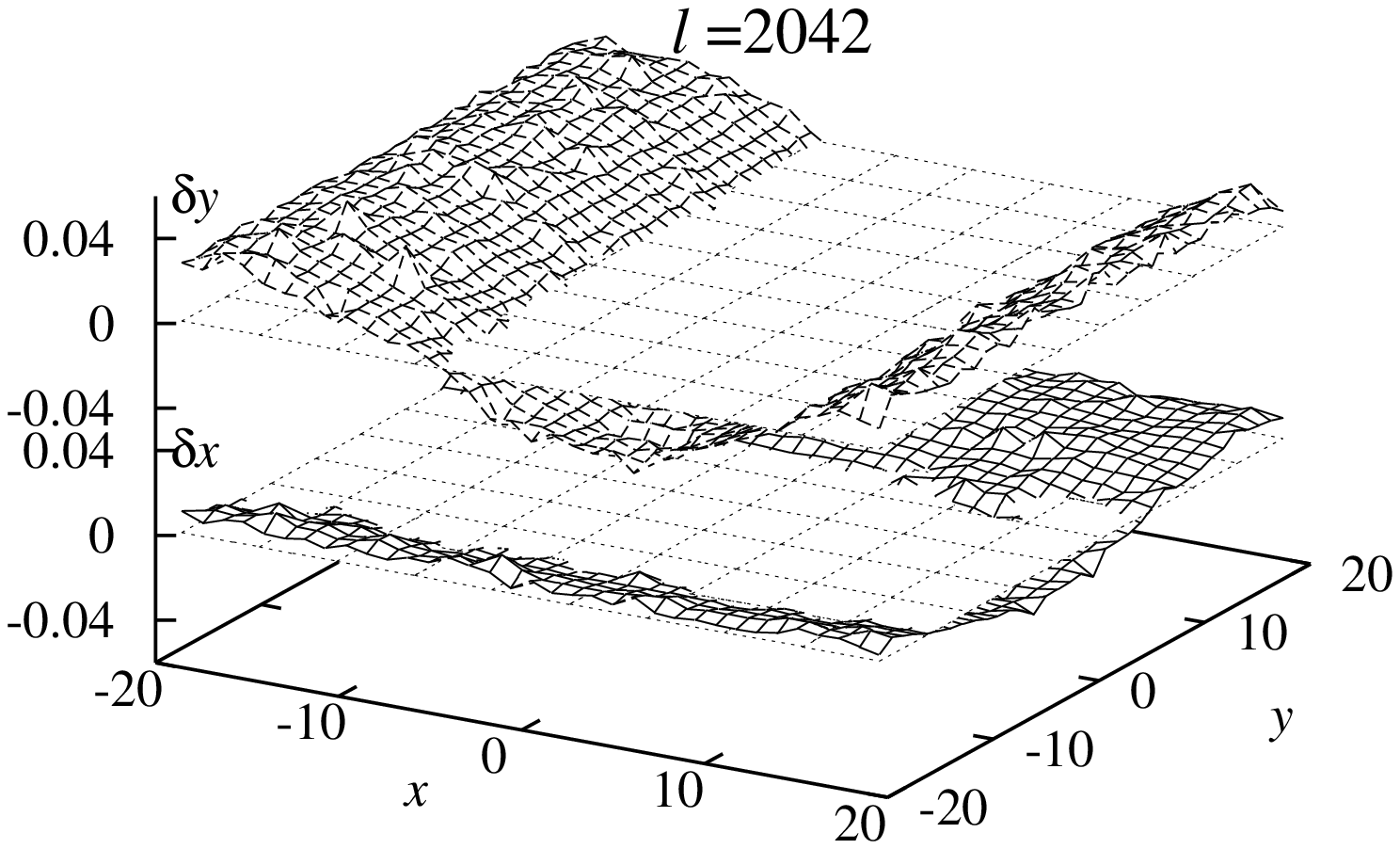}}
\hspace{1cm}
\centering{\includegraphics[width=5cm,clip=,angle=-90]{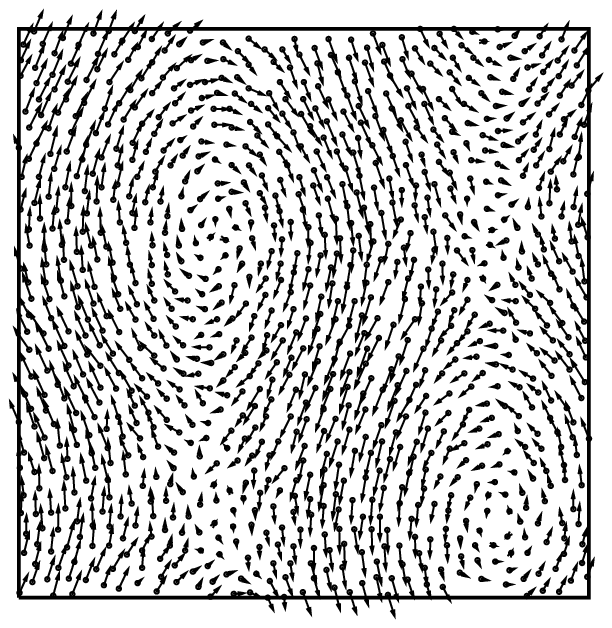}}
\caption{Transverse modes with the smallest-possible wave number for
$l=2045$, 2044, 2043 and 2042. The fluid consists of 1024 disks
with a density $\rho=0.7$. See the main text for details.}
\label{Fig5}
\end{figure}

    The first step consists of four degenerate exponents, for which
$l$ = 2045, 2044, 2043 and 2042. On the left-hand side of Fig. \ref{Fig5} 
we plot the positional perturbation components, $\delta x_i$ and 

$\delta y_i$, of the particles at their positions $(x_i,y_i)$ in the 
simulation box, where the surfaces have been linearly
interpolated on a periodic grid covering the simulation box.  
One finds periodic patterns like sines and cosines,
with a phase shift of $\pi/2$ between them. The wave vectors for 
$\delta x$ and $\delta y$  are orthogonal to the $y$ and $x$ axes, 
respectively, and we refer to these modes as {\em transverse}. 
Their wave number, $k \equiv |{\bf k}|$, is the smallest consistent 
with the periodic boundaries, $k_l =  2\pi/L$ for $2042 \le l \le 2044$.  
The same transverse modes, in another representation, are shown
on the right-hand side of Fig. \ref{Fig5}, where the
particle perturbations $(\delta x_i, \delta y_i)$ for all
particles $i$ are plotted as small arrows at the respective particle positions.
Transverse modes do not propagate.

  The next-larger exponents responsible for the second step in the inset
of Fig. \ref{Fig1} have multiplicity eight, 
$2034 \le l \le 2041$. Their modes are called {\em longitudinal}, since
the perturbations $\delta x$ and $\delta y $ are parallel to
their respective wave vectors with wave number $ 2\pi/L$. 
Longitudinal modes are propagating, with a phase velocity roughly equal to one 
third of the sound speed \cite{h99}. 

  Continuing up the steps, one encounters transverse modes for
$2030 \le l \le 2033$ with
multiplicity 4 and  wave vectors pointing along the diagonals of the
simulation box, $k = 2 \sqrt{2}  \pi/L$, followed by longitudinal
modes for $2030 \le l \le 2033$, with  multiplicity 8 and  wave vectors 
also along the diagonals, $k = 2 \sqrt{2} \pi/L$, and so on. 
In the general case of a rectangular simulation box, all modes look
like waves with $n_x$ nodes in $x$ direction and $n_y$ nodes in $y$ direction,
such that
\begin{equation}
k = 2 \pi \sqrt{\left(\frac{n_x}{L_x}\right)^2 +
                \left(\frac{n_y}{L_y}\right)^2  } \; .
\end{equation}
The case $n_x = n_y = 0$ corresponds to perturbation vectors in the central 
ma\-ni\-fold and has been treated in the previous section.
For all positive exponents the momentum perturbations $\delta p_x$ and
$\delta p_y$ are strictly parallel to the respective positional
perturbations  $\delta x$ and $\delta y$, and are therefore not included in
Fig. \ref{Fig5}.

\begin{figure}
\centering{\includegraphics[width=9cm,clip=,angle=-90]{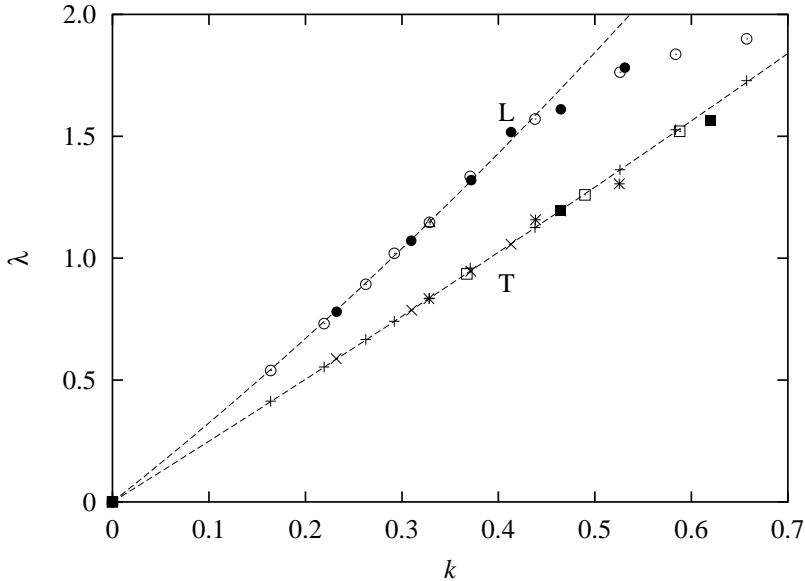}}
\caption{Dependence of the small positive Lyapunov exponents on
the wave number $k$ for various systems containing between 100 and 
1024 particles.  $L$ refers to longitudinal, and $T$ to transverse modes. 
The density of the hard-disk fluid is 0.7, the kinetic energy per particle,
$K/N$, is unity.}
\label{Fig6}
\end{figure}

    All modes may be classified either as transverse or longitudinal, 
where the dependence of the Lyapunov exponents on $k$ differs for each 
class.  This is shown in Fig. \ref{Fig6}, in which all small exponents,
plotted as a function of $k$, lie on two curves, one 
for the transverse (T) and one for the longitudinal (L) modes. To second
order in $k$, these ``dispersion relations'' may be approximated by
\begin{eqnarray}
   \lambda_T & = & 2.48(3) k + 0.23(7) k^2 + \mathcal{O}(k^3)   \label{lT}   \\
   \lambda_L & = & 3.13(7) k + 1.2(2) k^2 + \mathcal{O}(k^3) \;.\label{lL}
\end{eqnarray}
The standard deviations affect the last digit of the fit parameters and
are given in brackets. These second-order approximations are indicated as 
dashed lines in Fig. \ref{Fig6}.  It has been argued by Mareschal and
McNamara \cite{MMcN02} that the $k^2$ term is negligible. However,
the data, in particular the longitudinal exponents, are better
represented if this term is kept, although it is unessential for 
the discussion below. 

\subsection{Mode degeneracy}
\begin{figure}
\centering{\includegraphics[width=5cm,clip=]{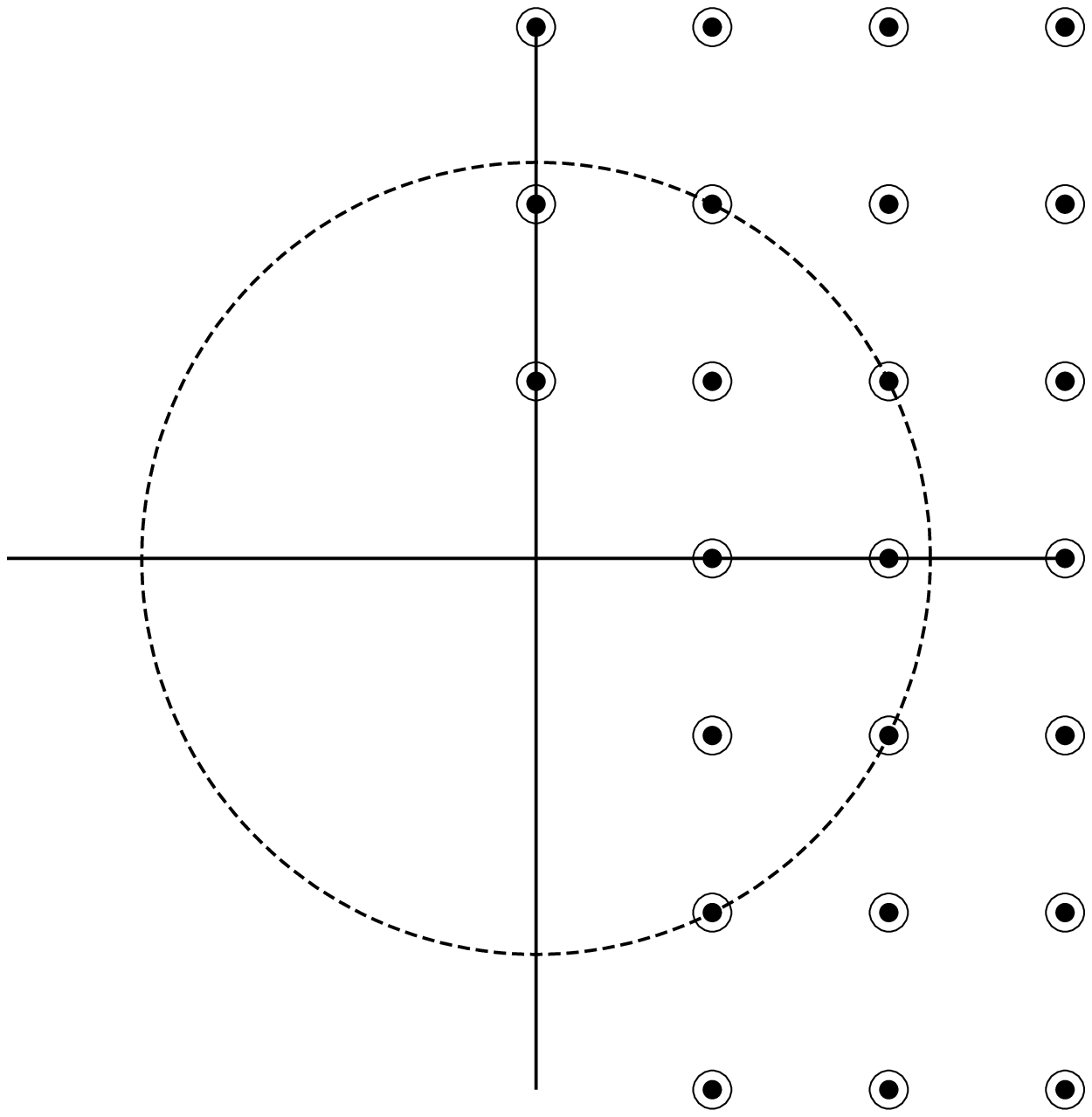}}
\hspace{1cm}
\centering{\includegraphics[width=5cm,clip=]{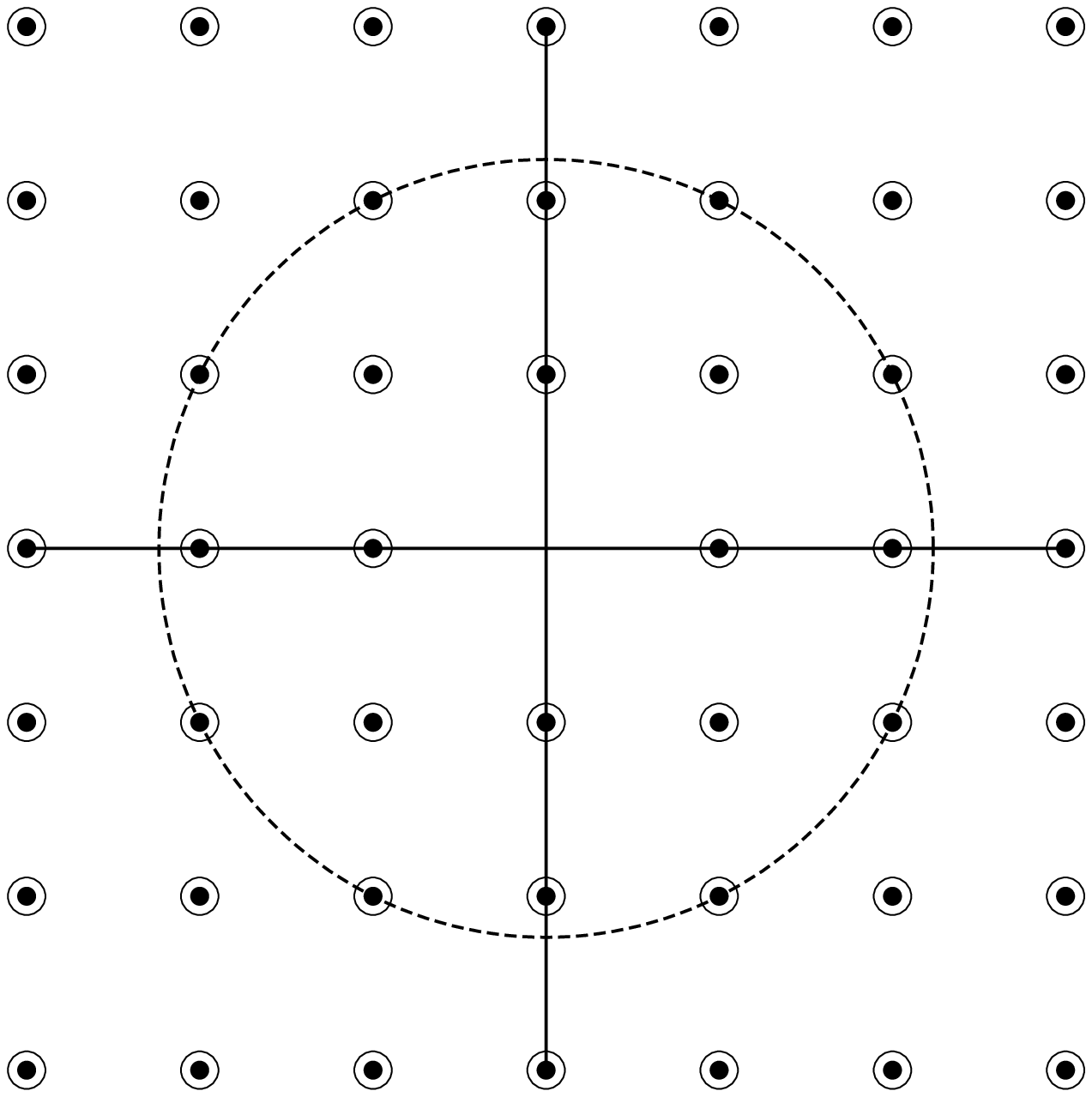}}
\caption{Reciprocal lattice with point spacing equal to $2 \pi/ L$,
for sine-like patterns (open circles) and cosine-like patterns (dots).  
Transverse modes do not require a sense of direction, and 
only one half plane is needed (left), whereas for 
longitudinal modes the full reciprocal lattice is required (right).
The circle connects allowed wave vectors with $k = 2 \sqrt{5} \pi/ L$.} 
\label{Fig7}
\end{figure}

     The degeneracy of the Lyapunov exponents is a consequence of the square 
symmetry of the periodic simulation box and the phase shift of $\pi/2$ between 
sine and cosine-like patterns, allowing for different {\em orthogonal} 
perturbations with the same $k$.  For the non-propagating {\em transverse} 
modes the sense of direction is not needed. It suffices to
consider all wave vectors in a half plane of the
reciprocal lattice, as on the left-hand side of Fig. \ref{Fig7}. 
\begin{figure}
\centering{\includegraphics[width=12cm,clip=]{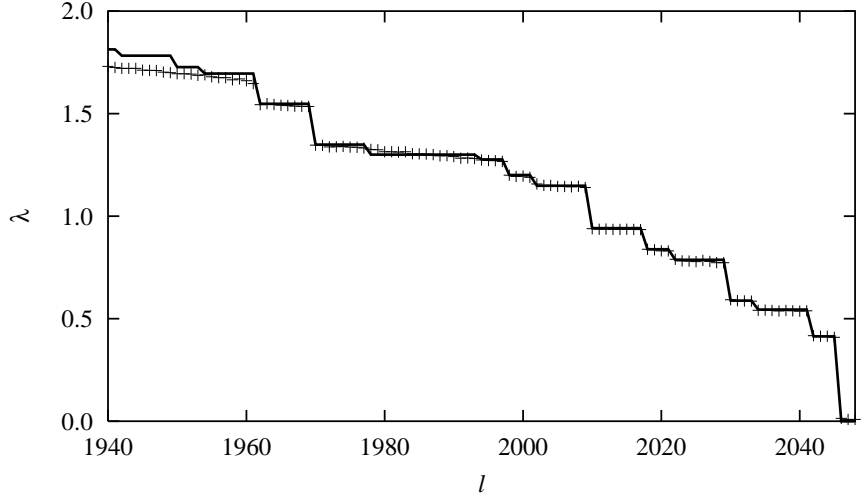}}
\caption{Small positive Lyapunov exponents for a 1024-disk
system: the crosses denote simulation results, the
smooth curve indicates the reconstruction to second order
in $k$ as described in the main text.  } 
\label{Fig8}
\end{figure}
The full and open circles refer to sine and cosine-like waves, respectively, 
which have to be counted separately. Degenerate modes belong to all lattice 
points on a circle with a specified  radius $k$. Their multiplicity is 
denoted by $M_k$. For the propagating 
{\em longitudinal} modes the full reciprocal lattice plane is needed as on 
the right-hand side of Fig. \ref{Fig7}.  As a consequence, the degeneracy 
of the longitudinal modes is always twice that of the transverse modes for a 
given $k$. With the circles in Fig. \ref{Fig7} we select degenerate modes 
with $k = 2 \sqrt{5} \pi /L$ as examples. The transverse modes have
multiplicity $8$ and  appear for $l = 2010, \dots , 2017$ in the 1024-disk 
spectrum of Fig \ref{Fig8}. The longitudinal modes 
have mutiplicity $16$ and are indexed by $l = 1978, \dots, 1993$.  Of course, 
the relative position of a degenerate group of exponents in the
spectrum is determined by the dispersion relations given above.

\subsection{Mode amplitudes and stability of transverse modes}
    Each mode with index $l$ is viewed here as a four-dimensional 
field on the periodic domain of the simulation box, and will be denoted
by\footnote{Sometimes it is 
advantageous to label the fields, $\mathcal{P}_l$, not by $l$ but by 
the wave-vector norm $k$ and an integer $m$ distinguishing
different modes within a degenerate group, $1\le m\le M_k$},
\begin{equation}
\mathcal{P}_l(x,y) = \{\delta x(x,y), \delta y(x,y), \delta p_x (x,y), 
\delta p_y (x,y)\}_l  \; .
\label{P}
\end{equation}
Its norm,
\begin{equation}
|\mathcal{P}_l(x,y)| = [(\delta x)^2+(\delta y)^2+(\delta p_x)^2+
                        (\delta p_y)^2]_l^{1/2} \; ,
\label{normam}
\end{equation}
is also a periodic function on the domain with amplitude $P_0^l$. 
Since $\delta {\bf \Gamma}_l$ is normalized, it follows that
\begin{equation}
\int_0^L\int_0^L |\mathcal{P}_l(x,y)|^2 \; dx dy = 1\;.
\label{norm}
\end{equation}

    As an example we consider again the transverse modes for $l=2N-3$ 
belonging to the first step in the spectrum. For $N = 1024$ such a mode is 
shown at the top of Fig. \ref{Fig5}. For any N it has the simple structure
\begin{eqnarray}
\mathcal{P}_{2N-3}(x,y)& = &\{(\delta x)_0 \sin(k y + \phi_y), \; 
                        (\delta y)_0 \sin(k x + \phi_x), \; \nonumber \\
                       &   &  (\delta p_x)_0 \sin(k y + \phi_y), \; 
                        (\delta p_y)_0 \sin(k x + \phi_x) \}_{2N-3} \; ,
\label{components}
\end{eqnarray}
where $k = 2\pi/L$. The phases $\phi_x$ and $\phi_y$ are equal for this 
particular example, but in a more general case they may differ by $\pi$. 

\begin{figure}
\centering{\includegraphics[width=12cm,clip=]{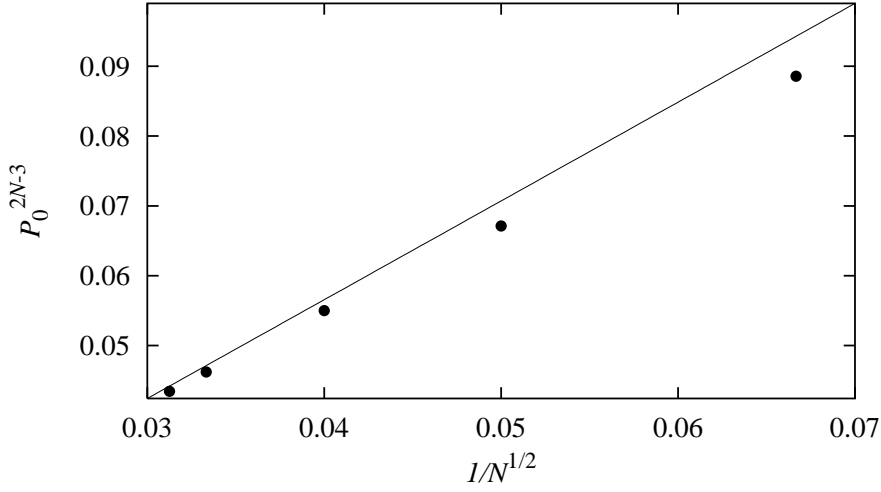}}
\caption{The dots indicate the  $1/\sqrt{N}$ dependence of the transverse-mode 
amplitude $P_0^{2N-3}$ corresponding to the first nonvanishing exponent
$\lambda_{2N-3}$. The smooth line is the  prediction of Eq. (\ref{amplitude}).}
\label{Fig9}
\end{figure}

  The mode amplitude may be expressed in terms of the component amplitudes 
according to
\begin{equation}
P_0^{2N-3} = [ (\delta x)_0^2 +  (\delta y)_0^2 +  (\delta p_x)_0^2 +
 (\delta p_y)_0^2 ]^{1/2} \; .
\end{equation}
The box-size and, hence, the particle-number dependence of this quantity 
follows immediately from the normalization condition Eq. (\ref{norm}):
\begin{equation}
              P_0^{2N-3} = \left(\frac{2}{N}\right)^{1/2} \; .
\label{amplitude}
\end{equation}
The amplitude vanishes in the thermodynamic limit \cite{Eckmann}.
In Fig. \ref{Fig9} the dependence of  $P_0^{2N-3}$ on $\sqrt{N} = L$
is shown. The dots are numerical results, where the mode amplitude
is obtained from fits of the components of Eq. (\ref{components}) to the data.
The smooth line is the result predicted by Eq. (\ref{amplitude}). 
Similar results are obtained for the other modes. The good agreement between
theory and experiment is an indication for the remarkable stability 
of the modes. 
      
    It should be pointed out that only the mode amplitude derived from
the vector norm, Eq. (\ref{normam}), displays the correct system-size 
dependence.
The amplitudes of the vector components do not. Even $(\delta x)_0$
and $(\delta y)_0$ do not agree in general. The amplitudes of the
field components depend on the initial conditions of the simulation,
but $P_0^l$ does not.

\subsection{Spectrum reconstruction near the thermodynamic limit}
\begin{figure}
\centering{\includegraphics[width=12cm,clip=]{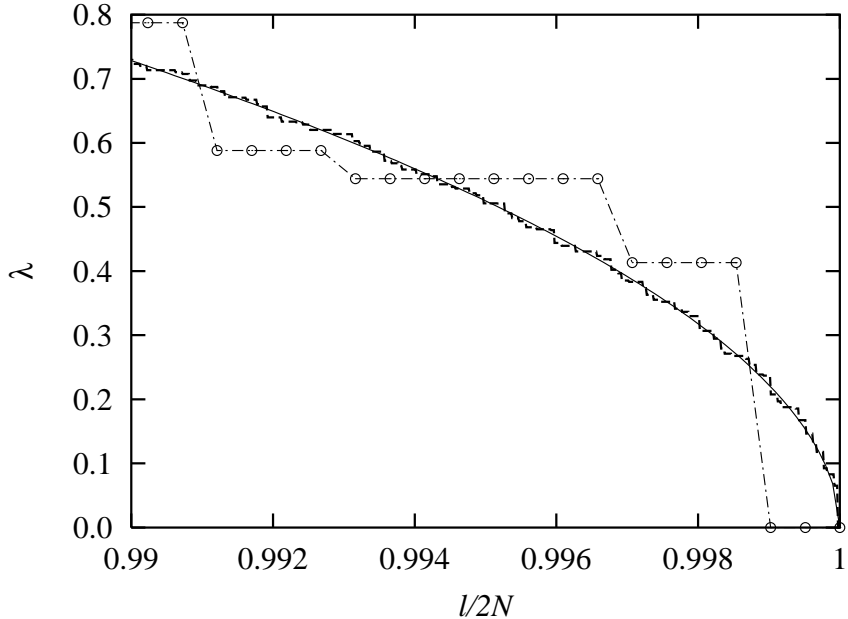}}
\caption{Part of a reconstructed Lyapunov spectrum for a fluid
of density 0.7,  consisting of 40,000 disks, is shown by the dashed line.
The smooth line is a fit of these points to a power law, Eq. (\ref{lfit}).
For comparison, the spectrum for 1024 disks is also shown by the points. 
On the abscissa a normalized index $l/2N$ is used.  } 
\label{Fig10}
\end{figure}

    If the dispersion relations are known for a fluid of a given density, 
the rules outlined above  allow a reconstruction of the most-interesting 
spectral range for small positive exponents. If the second-order
approximations of Eqs. (\ref{lT}) and (\ref{lL}) are used, the reconstructed
spectrum for the 1024-disk system is indicated in Fig. \ref{Fig8} by a smooth 
line. As expected, it agrees very well with the direct simulation results
for $l > 1962$ corresponding to small $k$.  Since Eqs. (\ref{lT})
and (\ref{lL})
apply to the $k \to 0$ limit, very large systems may be studied which are not
accessible by direct simulation, now and in the foreseeable future.
In Fig. \ref{Fig10} a reconstructed spectrum for a fluid containing 
$40,000$
disks with a density 0.7 is shown, which may be considered to be close to the
thermodynamic limit. Only a very small part of the spectrum is actually shown,
for which $k$ does not exceed 0.3.  We may fit a power law to these 
spectral points, 
\begin{equation}
\lambda_l = \alpha  \left[ 1 - \frac{l}{2N}\right]^{\beta} \;, \; 
     0.99 <   \frac{l}{2N} < 1 \; ,
\label{lfit}
\end{equation}
with $\alpha = 7.81 \pm 0.04$ and $\beta = 0.52 \pm 0.01$.\footnote{
We note that for a simple model of a two-dimensional solid the number of 
vibrational modes $d l$ between wave numbers $k$ and $k + d k$ is 
proportional to $k$.  Integrating this relation and assuming linear dispersion 
relations for the Lyapunov exponents we obtain $\lambda \sim l^{1/2}$ 
\cite{ph88}. This power closely agrees with our value for $\beta$.}
The slope of the spectrum near the intersection with the abscissa,
\begin{equation}
 \lim_{l/2N \to 1} \frac{d\lambda_l}{d(l/2N)} \sim \lim_{l/2N \to 1} \left[1 - 
         \frac{l}{2N}\right]^{-0.48}
\label{dlfit}
\end{equation}
diverges for $l/2N \to 1$. Since the small positive exponents 
approach zero linearly with $k$ as expressed by the dispersion relations 
above, no gap appears in the spectrum. No lower positive bound exists for the
positive Lyapunov exponents. An analogous statement for the
negative exponents of the spectrum holds due to the conjugate pairing
symmetry of the spectrum.

     We have also extended these considerations to quasi-one-dimensional 
systems \cite{fp03,fdiplom}, for which the aspect ratio of the simulation 
box, $A \equiv L_y/L_x $, converges to zero in the large-particle limit.
This is achieved by fixing the box size $L_y$ in $y$ direction, typically
a few particle diameters,  and letting $L_x$ become large such that the
particle density remains constant. The Lyapunov modes are more obvious
in this case, since the reciprocal lattice consists of 
equidistant points on the abscissa, allowing only multiplicity  two for the 
transverse modes, and four for the longitudinal. The limiting spectral slope 
near the intersection with the abscissa remains finite \cite{fp03}.
Very recently, Taniguchi and Morriss \cite{TM02b} have also studied other than 
periodic  boundary conditions for very narrow quasi-one-dimensional systems, 
for which the particles cannot cross. Equivalent modes are observed.

     So far we have concentrated on the positive branch of the 
Lyapunov spectrum,
for which the components of the momentum perturbations are parallel to the
respective perturbation components for the positions. This is a 
consequence of the linearity of the motion equations in tangent space and the
positivity of the exponents. For the negative branch
of the spectrum the momentum components are antiparallel to the position
components of the perturbations. All our statements concerning the
multiplicity and polarization of the modes carry over also to this case.

\section{A puzzle provided by the soft potential}
\label{Sec6}

   What happens if the interparticle potential is soft? In Refs. 
\cite{hpfdz,pf02} a comparison was made between soft and hard-disk systems and,
surprisingly, no modes could be clearly identified.  Here we extend this work 
by considering two-dimensional soft fluids with a 
Weeks-Chandler-Anderson (WCA) interaction potential,
\begin{equation}
\phi(r)=\left\{
    \begin{array}{ll}
        4\epsilon[(\sigma/r)^{12}-(\sigma/r)^{6}]+\epsilon, & r<2^{1/6}\sigma\\
        0, & r\ge 2^{1/6}\sigma \; .
    \end{array}\right. 
\end{equation}
We use units for which
the particle mass $m$, the particle diameter $\sigma$, and the energy parameter
$\epsilon$ are unity. Furthermore, we consider a
thermodynamic state with a total energy per particle, $E/N$, also equal to 
unity.  Since there is a potential energy in this case, a
comparison between soft and hard disks is done for 
equal temperatures $T \equiv K/N$ (and not for equal energy), where
Boltzmann's constant is also set to unity.  All hard-disk Lyapunov exponents
quoted in this section are therefore rescaled by a factor
$\sqrt{(K/N)_{WCA}/(K/N)_{HD}}$, where $K$ is the total kinetic energy.
Since the soft-potential spectra are computationally more expensive, we 
restrict our examples to $N \le 100$.

\begin{figure}
\centering{\includegraphics[width=8cm,clip=,angle=-90]{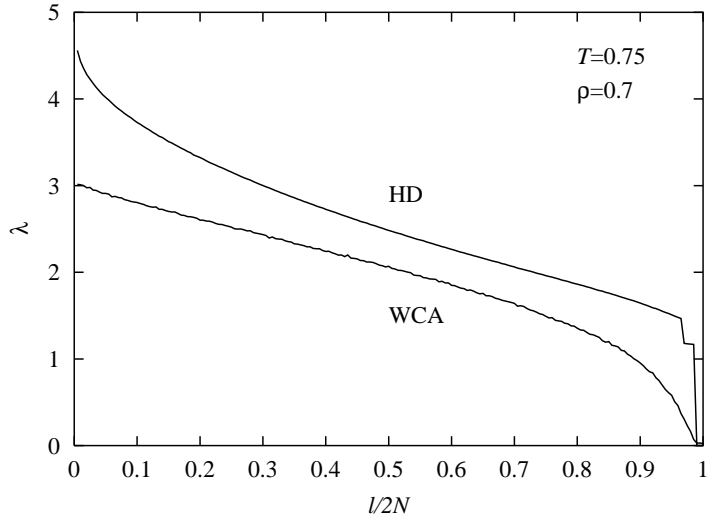}}
\caption{Lyapunov spectra for 100 soft (WCA) and hard (HD)
disks at a density $\rho = 0.7$ and a temperature
$T = 0.75$. The simulation box is a square with periodic boundary
conditions}
\label{Fig11}
\end{figure}

\begin{figure}
\centering{\includegraphics[width=8cm,clip=,angle=-90]{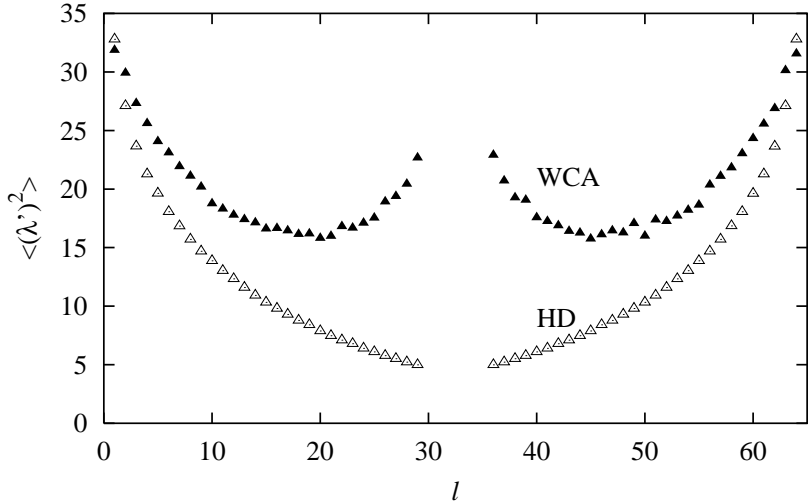}}
\caption{Second moments of the time-dependent local exponents
for a planar 16-particle system. The labels WCA and HD refer to the
Weeks-Chandler-Anderson and hard disk models, respectively.
The re-orthonormalization time $\Delta t_{GS} = 0.075$,
the temperature $T= 0.75$, and the density $ \rho = 0.7$.}
\label{Fig12}
\end{figure}

   The  spectra for the soft- and hard-disk fluids in Fig. \ref{Fig11}
are surprisingly different, both in shape and in absolute 
value.  Most conspicuously, no steps are observed in the WCA case,
whereas for hard disks at least the lowest step is well developed. 
This comparison provides us with the following facts \cite{pf02,fp03}: \\
1) The overall shape of the WCA spectrum may  be approximated by a 
   power law similar to Eq. (\ref{lfit}), but with a power of the order of 0.4,
   (which depends slightly on the fitting range). A similar behavior has been
   found for three-dimensional WCA fluids, and has been interpreted there
   in terms of a simple Debye model for the distribution of vibrational
   frequencies in solids, anticipating the Lyapunov modes.
   For details we refer to Ref. \cite{ph88}, although the systems there are too
   small for a meaningful comparison with the hard-disk fluids.  \\
2) If the WCA potential is made progressively steeper, its spectrum
   converges towards that of hard disks, and the step structure
   starts to re-appear \cite{fp03}. 
   This convergence, however, is slow, in particular for the largest 
   exponent $\lambda_1$.
3) The perturbation vector associated with $\lambda_1$ is also localized
   for soft disks, and this localization persists in the thermodynamic
   limit. An instantaneous snapshot of such a perturbation for 100 WCA 
   particles looks very similar to Fig. \ref{Fig2}. 
   Also the localization measure $C_{1,\Theta}$ introduced in Section 
   \ref{Sec3} is found to obey a power law with a negative 
   power of $N$ \cite{fdiplom}: $C_{1, \Theta = 0.92} = 0.31 N^{-0.10}$
   for a WCA fluid with $\rho = 0.7$ and $T=0.75$.
   Only a few particles, localized in space,  provide the main
   contributions to the perturbation vector $\delta {\bf \Gamma}_1$. 
   However, there is a very sensitive dependence of the details 
   on the interparticle potential. \\
4) As we suspected from the absence of a step structure, no
   Lyapunov modes could be reliably identified, either visually or
   by Fourier-transformation techniques. Although it is 
   possible to see patterns resembling transverse modes for the smallest
   exponent, these patterns are transient and too unstable
   to survive time averaging \cite{hpfdz}. 
   This observation is consistent with large fluctuations of the local
   time-dependent exponents $\lambda'_l$, from which the Lyapunov
   exponents are obtained by time averaging:
\begin{equation}
\lambda_l = \lim_{ \tau \to \infty}\frac{1}{\tau}
\int_{0}^{\tau} {\lambda'}_l ({\bf \Gamma}(t)) dt \; .
\end{equation}
$ {\lambda'}_l({\bf \Gamma})$ depends on the state ${\bf \Gamma}(t)$
the system occupies in phase space at time $t$ and, of course, on
a long-enough part of the trajectory terminating at this point. They may be
estimated from
\begin{equation}
{\lambda'}_l({\bf \Gamma}(t)) =\frac{1}{\Delta t_{GS}}
 \ln \frac{|\delta{\bf \Gamma}_l({\bf \Gamma}(t+ \Delta t_{GS})|}
          {|\delta{\bf \Gamma}_l({\bf \Gamma}(t)|},
\label{local}
\end{equation}
where $t$ and $t+\Delta t_{GS}$ refer to times immediately
after consecutive Gram-Schmidt re-orthonormalization steps.
In Fig. \ref{Fig12} the 
second moments of these quantities, $\{\langle  {\lambda'}_l^2 \rangle\},\; 
l=1,\dots,4N\}$,
time averaged along the phase trajectory, are shown for 16-particle WCA and 
hard-disk systems, where a re-ortho\-nor\-ma\-li\-za\-tion interval 0.075 is 
used\footnote{The computation of the second moments
$\langle  {\lambda'}_l^2 \rangle$ for hard disks requires some care.
For small re-orthonormalization intervals $ \Delta t_{GS}$, 
the variance $\langle  {\lambda'}_l^2 \rangle - \langle  
{\lambda'}_l \rangle^2$ varies with $1/ \Delta t_{GS}$ and diverges
for $ \Delta t_{GS}\to 0$, but the shape of the
fluctuation spectrum is hardly affected by the size of $\Delta t_{GS}$.}.  
$l=1$ refers to the maximum, and $l = 64$ to the most-negative exponent.
The points for $30 \le l \le 35$ correspond to the 6 vanishing exponents
and are not shown.  We infer from this figure that for the soft WCA particles
the fluctuations of the local exponents become very large
for the Lyapunov exponents describing relatively-weak instabilities with
near-zero growth rates, $l \to 2N$. For the hard disk system, however,
the relative significance of the fluctuations becomes minimal in this regime.

\section{Outlook}
\label{Sec7}

   Soon after the discovery of the Lyapunov modes, Eckmann and Gat 
\cite{Eckmann} were the first to provide theoretical arguments for the
existence of hydrodynamic-like Lyapunov modes. They were derived
from the translation invariance of the systems, and were based on an  
evolution matrix in tangent space modeled as a product of independent random 
matrices. No accompanying real space dynamics exists for this model.  
Most recently, this
theory has been improved and made more realistic by Eckmann and Zabey
\cite{Zabey}.  In another approach, McNamara and Mareschal \cite{McNM01}
isolate the six hydrodynamic fields related to the invariants of the binary
particle collisions and the vanishing exponents, and use
a generalized Enskog theory to derive hydrodynamic evolution equations for
these fields.  Their solutions are the Lyapunov modes. In a more detailed
extension of this work a generalized Boltzmann equation is used for the
derivation of the hydrodynamic evolution equations \cite{MMcN02},
which restricts the quantitative aspects of the theory to small densities.
A related approach was taken by de Wijn and van Beijeren \cite{deWvB},
who have pointed out that the modes should be interpreted as Goldstone modes:
The breaking of a continuous symmetry -- translational invariance in our case --
implies the existence of hydrodynamic-like ``soft'' modes and long-ranged
correlations.  
Finally, Taniguchi, Dettmann, and Morriss have approached the problem
using periodic orbit theory \cite{TDM02} and master equations \cite{TM02}.

   We have mentioned in Section \ref{Sec6} that for small positive
exponents the Lyapunov spectrum is well described by a power-law close to the
thermodynamic limit, and that no positive lower bound exists in this case. 
This has raised questions about the numerical accuracy and convergence of our 
results. With respect to the former, we have verified that quadruple precision 
(128 bits) arithmetic  gives the same results as the usual double-precision 
arithmetic. To answer the question of convergence, we have perturbed an
already well-oriented set of orthonormal tangent vectors for a 
well-relaxed system and have ascertained that the unperturbed and 
perturbed sets of vectors become identical again after a certain
transient time. The relaxation time the tangent vectors require to reach their
proper orientations has been studied in Ref. \cite{dhp02}.

   The spatial localization of the fastest-growing perturbation
described by the maximum exponent is a  very general result. It is also found
for particle systems interacting {\em via} soft potentials \cite{pf02}, 
although the details of the
power-law dependence of the localization measure $C_{1,\Theta}$ depend
sensitively on the interaction potential.  Also continuous
systems in nonequilibrium stationary states exhibiting Rayleigh-B\'enard
convection have this localization property \cite{egolf}. There it is found that
the localized active zones contributing most to the perturbation 
vector for large positive Lyapunov exponents are involved in the creation
and annihilation of defects of the roll pattern, resulting in the
breaking or reconnection of Rayleigh-B\'enard rolls \cite{egolf}.

\section*{Acknowledgements} We enjoyed and  acknowledge many useful 
discussions with Predrag Cvitanovi\'c, 
Christoph Dellago, Astrid de Wijn, Bob Dorfman, Jean-Pierre Eckmann,  
Denis Evans, Rainer Klages, Joel Lebowitz, Ljubomir Milanovi\'c, Gary Morriss, 
David Mukamel,
G\"unter Radons, David Ruelle, Ya. G. Sinai, Tooru Taniguchi, Walter Thirring, 
Henk van Beijeren, J\"urgen Vollmer, Emmanuel Zabey, and participants of the 
workshop {\em  Microscopic Chaos and Transport in Many-Particle Systems},
August 2002, in Dresden. This work was supported by the
Austrian Fonds zur F\"orderung der wissenschaftlichen Forschung,
Grants P11428-PHY and P15348-PHY.


\begin{thebibliography}{00}




\bibitem{ph88} H. A. Posch and W. G. Hoover,
        Phys. Rev. A {\bf 38},  473 (1988).

\bibitem{PH89} H. A. Posch and W. G. Hoover,
        Phys. Rev. A {\bf 39}, 2175 (1989).

\bibitem{HOO} Wm. G. Hoover, {\em Time Reversibility, Computer Simulation,
        and Chaos}, ( World Scientific, Singapore, 1999).

\bibitem{em90} D. J. Evans and G. P. Morriss, {\em Statistical mechanics
        of nonequilibrium liquids}, Academic Press, London, 1990.

\bibitem{GASP} P. Gaspard, {\em Chaos, Scattering, and Statistical Mechanics},
        (Cambridge University Press, 1998).

\bibitem{DO99} J. R. Dorfman, {\em An Introduction to Chaos in 
        Nonequilibrium Statistical Mechanics}, 
        (Cambridge University Press, 1999).

\bibitem{ph00} H. A. Posch and R. Hirschl, ``Simulation of Billiards and of
      Hard-Body Fluids'', p. 269 - 310, 
      in {\em Hard Ball Systems and the Lorenz Gas},
      edited by D. Szasz, Encyclopedia of the mathematical sciences
      {\bf 101}, Springer Verlag, Berlin (2000).
 
\bibitem{h99} R. Hirschl, ``Computer simulation of hard-disk and hard-sphere 
      systems: Lyapunov modes and stochastic color conductivity'',
      Master's thesis, University of Vienna, 1999.
 
\bibitem{pf02} H. A. Posch and Ch. Forster, ``Lyapunov instability and 
      collective tangent space dynamics of fluids'', Lecture Notes on
      Computer Science {\bf 2331}, {\em Computational Science - ICCS 2002},
      vol. 3, p.1170,
      edited by P.M.A. Sloot, C.J.K. Tan, J.J. Dongarra, and A.G. Hoekstra,
      Springer, Berlin, 2002.  

\bibitem{hpfdz} Wm. G. Hoover, H. A. Posch, Ch. Forster, Ch. Dellago, 
       and M. Zhou, J. Stat. Phys. {\bf 109}, 765 (2002).

\bibitem{mph98a} Lj. Milanovi\'c, H.A. Posch, and Wm. G. Hoover, 
        CHAOS {\bf 8}, 455 - 461 (1998).

\bibitem{mph98b} Lj. Milanovi\'c, H.A. Posch, and Wm. G. Hoover,
        Molec. Phys., {\bf 95}, 281 - 287 (1998).

\bibitem{mp02} Lj. Milanovi\'c and H. A. Posch, 
        J. Molec. Liquids, {\bf 96-97}, 221 - 244 (2002).

\bibitem{m01} Lj. Milanovi\'c, ``Dynamical instability of two-dimensional
        molecular fluids: hard dumbbells'', Ph.D. thesis, University of
        Vienna, 2001.

\bibitem{BERNAL} J. D. Bernal and S. V. King, in
        {\em Physics of Simple Liquids}, H. N. V. Temperley,
        J. S. Rowlinson, and G. S. Rushbrooke eds., page 231,
        (North-Holland, Amsterdam, 1968).

\bibitem{b3} B. J. Alder and T. E. Wainwright, Sci. Am. \bf 201\rm(4),
         113 (1959).

\bibitem{EIN}   T. Einwohner and B. J. Alder, J. Chem. Phys. {\bf 49},
         1458 (1968).

\bibitem{Hansen} J.-P. Hansen and I. R. McDonald, {\em Theory of simple 
        liquids}, (Academic Press, London, 1991).

\bibitem{Reed} T. M. Reed and K. E. Gubbins, {\em Applied Statistical
        Mechanics}, (McGraw Hill, Tokyo, 1973).

\bibitem{Eckmann} J.-P. Eckmann and O. Gat, J. Stat. Phys. {\bf 98}, 775 (2000).

\bibitem{Zabey} J.-P. Eckmann and E. Zabey, private communication, and 
       {\em International Workshop and Seminar
       on Microscopic Chaos and Transport in Many-Particle Systems},
       (Max Planck-Institut f\"ur Physik komplexer Systeme, Dresden, 2002), 
       unpublished.
        
\bibitem{McNM01} S. McNamara and M. Mareschal, Phys. Rev. E {\bf 64},
        051103 (2001). 

\bibitem{MMcN02}  M. Mareschal and S. McNamara,
       {\em International Workshop and Seminar
       on Microscopic Chaos and Transport in Many-Particle Systems},
       (Max Planck-Institut f\"ur Physik komplexer Systeme, Dresden, 2002), 
       this issue.

\bibitem{deWvB} A. de Wijn and H. van Beijeren, private communication, and
       {\em International Workshop and Seminar
       on Microscopic Chaos and Transport in Many-Particle Systems},
       (Max Planck-Institut f\"ur Physik komplexer Systeme, Dresden, 2002), 
       this issue.

\bibitem{TDM02} T. Taniguchi, C. P. Dettmann, and G. P. Morriss,
       J. Stat. Phys. {\bf 109}, 747 (2002).

\bibitem{TM02} T. Taniguchi  and G. P. Morriss,
       Phys. Rev. E {\bf 65}, 056202 (2002).

\bibitem{DPH96} Ch. Dellago, H.A. Posch, and Wm. G. Hoover, Phys. Rev. E
        {\bf 53}, 1485 (1996).

\bibitem{HBP98} Wm. G. Hoover, K. Boercker, and H. A. Posch,
        Phys. Rev. E {\bf 57}, 3911 (1998).

\bibitem{oseledec} V. I. Oseledec,
        Trans. Mosc. Math. Soc. {\bf 19}, 197 (1968).

\bibitem{eckmann} J.-P. Eckmann and D. Ruelle, Rev. Mod. Phys. { \bf 57},
        617 (1985).

\bibitem{DM96} C. P. Dettmann and G. P. Morriss, 
        Phys.Rev. E {\bf 53}, R5541 (1996).

\bibitem{Ruelle}
        D. Ruelle, J. Stat. Phys. {\bf 95}, 393 (1999).

\bibitem{ECM90} D. J. Evans, E. G. D. Cohen, and G. P. Morriss,
        Phys. Rev. A {bf 42}, 5990 (1990).

\bibitem{Benettin}
        G. Benettin, L. Galgani, A. Giorgilli, and J. M. Strelcyn,
        Meccanica {\bf 15}, 9 (1980).

\bibitem{Shimada}
        I. Shimada and T. Nagashima, Proc. Theor. Phys. {\bf 61}, 1605 (1979).

\bibitem{fp03} Ch. Forster and H. A. Posch, unpublished.

\bibitem{fdiplom} Ch. Forster, ``Lyapunov instability of two-dimensional
         fluids'', Master's Thesis, University of Vienna, 2002.

\bibitem{TM02b} T. Taniguchi and G. P. Morriss, preprint.

\bibitem{dhp02} Ch. Dellago, Wm. G. Hoover, and H. A. Posch,
         Phys. Rev. E {\bf 65}, 056216 (2002).

\bibitem{egolf} D. A. Egolf, I. V. Melnikov, W. Pesch, and R. E. Ecke,
         Nature {\bf 404}, 733 (2000).
\end{thebibliography}
\end{document}